\documentclass[letterpaper,11pt]{article}

\usepackage{amsmath,amssymb,textcomp,graphicx,array,natbib,soul, bm} 
\usepackage{fullpage,rotate}
\usepackage[small,bf,up]{caption}
\usepackage{snapshot}
\usepackage{xcolor}

\usepackage{setspace}
\begin{document}
\noindent
{\LARGE \sf  Calibrating a Stochastic Agent Based Model Using
 Quantile-based Emulation
}\\[0.5cm]
Arindam Fadikar, NDSSL, Biocomplexity Institute of Virginia Tech\\
Dave Higdon, Social Decisions Analytics Lab, Biocomplexity Institute of Virginia Tech\\
Jiangzhuo Chen, NDSSL, Biocomplexity Institute of Virginia Tech\\
Brian Lewis, NDSSL, Biocomplexity Institute of Virginia Tech\\
Srini Venkatramanan, NDSSL, Biocomplexity Institute of Virginia Tech\\
Madhav Marathe, NDSSL, Biocomplexity Institute of Virginia Tech\\

\graphicspath{{./}{Figs/}}

\noindent
{\small
 In a number of cases, the Quantile Gaussian Process (QGP) has proven effective in 
emulating stochastic, univariate computer model output \citep{plumlee2014building}.
In this paper, we develop an approach that uses this emulation approach within a Bayesian
model calibration framework to calibrate an agent-based model of an epidemic.  In addition,
this approach is extended to handle the multivariate nature of the model output, which gives
a time series of the count of infected individuals.  The basic modeling approach is adapted
from \citet{higdon2008cmc}, using a basis representation to capture the multivariate
model output.  The approach is motivated with an example taken from the 2015 
Ebola Challenge workshop which simulated an ebola epidemic to evaluate methodology.
\\[.2cm]
}


\section{Introduction}
\label{sec:intro}
This  paper develops statistical methodology to calibrate unknown parameters of a multivariate stochastic computer model, and to produce prediction uncertainty in the resulting model-based forecasts. For the purposes of this paper, we develop and demonstrate this methodology by combining epidemic observations with a stochastic, agent-based model (ABM) of the disease epidemic.  More details regarding this ABM are given in Section \ref{sec:abm}.

In general, ABMs are computationally demanding and stochastic, complicating their use within formal statistical inference.  Computational models with stochastic output have been used to carry out sensitivity analyses \citep{marrel2012global}, computer model emulation \citep{plumlee2014building}, 
optimization \citep{kleijnen2009kriging}, and calibration \citep{henderson2009bayesian,andrianakis2015bayesian}.  By the term {\em Bayesian model calibration}, we mean combining a collection of output resulting from computer model runs with physical observations of the system to estimate the posterior distribution of the input model parameters, as well as estimating the error between model and reality so that a posterior predictive distribution for future outcomes, or outcomes under different conditions is obtained.

Purely sampling-based, or Monte Carlo-based approaches for sensitivity analysis \citep{sobol2001global,saltelli2008global} or inference \citep{cornuet2008inferring,flury2011bayesian} carry out their inference by making direct and repeated use of the stochastic model output. In cases where running the computational model is too demanding for Monte Carlo-based estimation, emulation -- modeling the input-output response surface -- has proven to be an effective approach.  

In some applications, one only requires the mean response as a function of the input parameters \citep{kleijnen2009kriging,lawrence2010coyote,andrianakis2017efficient}. In such cases the collection of realizations can be averaged, preprocessed, or accounted for in the estimation scheme for the emulator.
In other cases the distribution of computer model outcomes (at a given input setting) can be described by its mean and variance.  One strategy is to model the output distribution as normal, conditional on the mean and variance and use a Gaussian process (GP) to model the mean and variance as a function of the inputs \citep{henderson2009bayesian,marrel2012global}. Adapting similar strategies to non-Gaussian distributions has also been carried out -- see \citet{reich2012variable} for an example of modeling the exposure distribution as a mixture of normals, using a formulation that depends on space, time, and individual-level covariates.

An alternative is the quantile kriging approach described in \citet{plumlee2014building}.  Here quantiles of the output distribution are estimated from the collection of realizations produced  at each input setting.  A GP is then fit to the augmented input space consisting of the original inputs combined with the quantile estimates.  Once fit to the model output, this results in a GP whose realizations give continuous, univariate distributions at each input setting.

In this paper we extend the quantile-kriging emulator to handle functional, stochastic computer model output.  We also embed this new emulation approach within a Bayesian computer model calibration framework 
\citep{kenn:ohag:2001,higdon2008cmc} to take on a challenging problem in characterizing epidemic behavior, using an ABM that models interaction and contact at the individual level.
This statistical modeling is implemented in the software of GPMSA \citep{gattuq:2016}, taking advantage of thoughtful preprocessing of the epidemic observations and the simulation output.

\section{The Ebola Challenge Problem}
\label{sec:ebolachallenge}

The 2015 ebola challenge problem 
({\tt ebola-challenge.org}) was organized by the Research and Policy for Infectious Disease Dynamics (RAPIDD) program
at the National Institutes for Health (NIH).  Its goal is to motivate the development, assessment, and comparison of different approaches for modeling, predicting, and managing infectious disease epidemics.

The challenge involved multiple scenarios for the nation of Liberia, each simulating an epidemic patterned after the Ebola outbreak of 2014.  Here we focus on Scenario 5 where weekly nationwide counts of ebola cases were reported at regular intervals.  For motivating our methodological approach, we take the observations to be the number of cases up to week 20, and focus on prediction using this data along with the ABM constructed for this challenge.

\subsection{The Agent-Based Epidemic Model for Liberia}
\label{sec:abm}

This model was developed to simulate the Ebola epidemic of 2014, and later updated to participate in the 2015 Ebola challenge ({\tt ebola-challenge.org}) organized by the Research and Policy for Infectious Disease Dynamics (RAPIDD) program
at the National Institutes for Health (NIH).
The components of the model include a realistic synthetic population, a social contact network induced by movement and interaction of individuals, and a disease model.

\begin{figure}[!ht]
\centering
\includegraphics[width=0.9\textwidth]{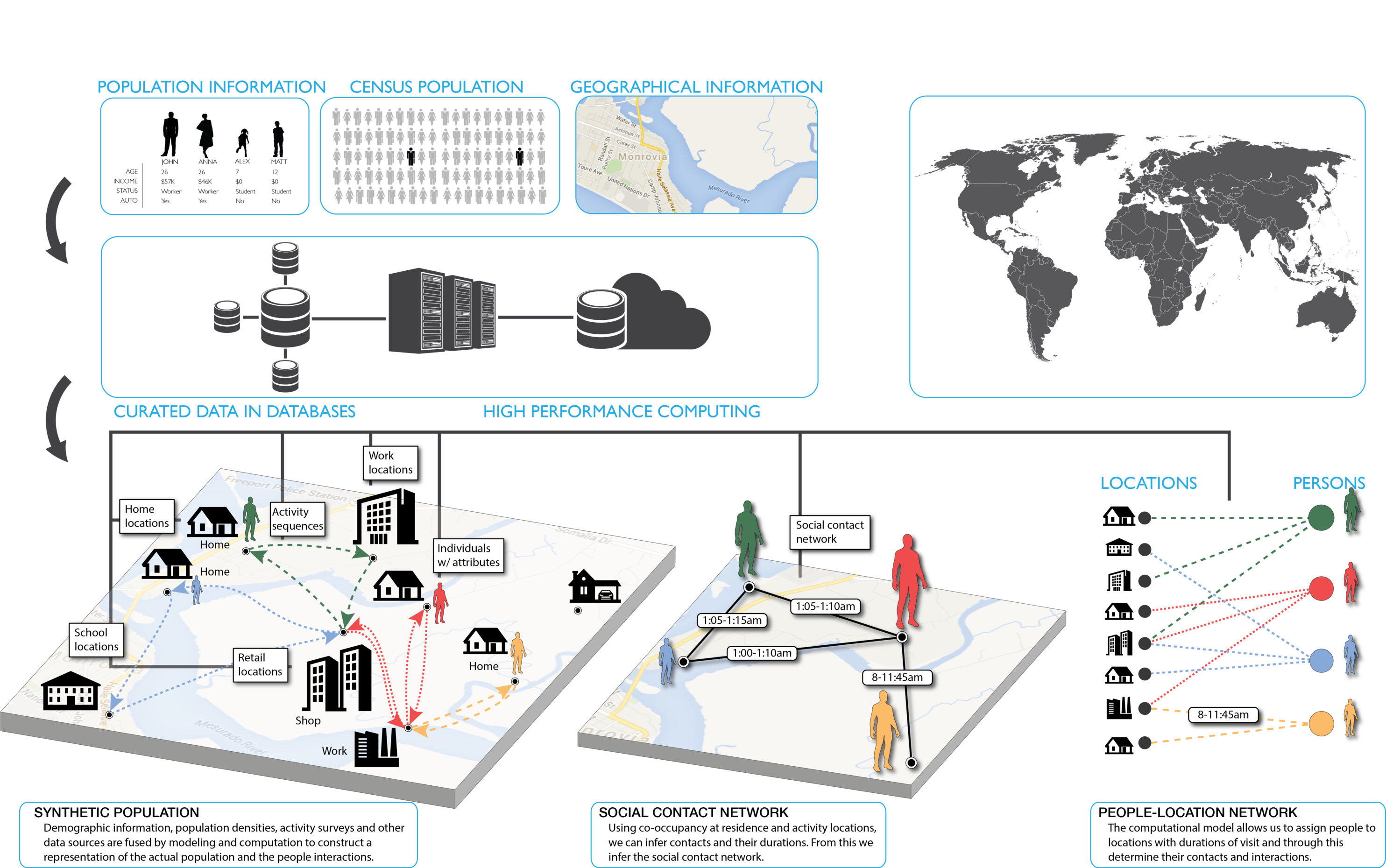}
\caption{A detailed view of synthetic population generation process. It consists of generating (a) synthetic population of individuals with explicit spatial  informations, (b) a mapping between the individuals and activity locations, and (c) an edge-weighted social contact network for disease propagation. (image courtesy: Henning Mortveit)}
\label{fig:synth_pop}
\end{figure}

The synthetic population is derived from a combination of survey data, LandScan satellite images, and demographic information taken from a similar country (Nigeria).  The resulting population (available at { \tt www.bi.vt.edu/ndssl/projects/ebola}) represents each individual in the country; each as part of a household or family, living at a specified location in Liberia.  As seen in Figure \ref{fig:synth_pop}, individuals move between home, work, and school locations according to realistic, individualized activity schedules that detail individual movement between these different locations over the course of the day. 

As individuals move, they come in contact with one another, inducing a time-varying contact network of individuals and locations, making it possible for the contagious disease to spread from one individual to another. 
\begin{table}[!b]
\caption{Epidemic model parameters and their lower and upper bounds.}
\small
\label{tab:params}
\centerline{
\begin{tabular}{cccc}
\hline\hline\noalign{\smallskip}
parameter & description & lower & upper \\
\noalign{\smallskip} \hline \noalign{\smallskip}
$\theta_1$ & transmissibility &  $3\times10^{-5}$ & $8\times10^{-5}$ \\
$\theta_2$ & initial infected number & 1 & 20 \\
$\theta_3$ & hospital intervention delay & 2 & 10 \\
$\theta_4$ & hospital intervention efficacy & 0.1 & 0.8 \\
$\theta_5$ & intervention travel reduction & 
0 & 2 \\
\noalign{\smallskip} \hline\hline
\end{tabular}
} 
\end{table}%
The probability that the disease is transmitted from an infected individual to a susceptible individual depends on the duration of the contact, as well as  other parameters that govern the agent-based model.  The transmissibility parameter $\theta_1$ modifies this transmission probability.  After contact, an individual may move to an exposed state, followed by the infected state.  Once infected an individual may infect others until he/she is removed from the population (by staying home, seeking treatment, or possibly dying).  The parameters $\theta_3,\ldots,\theta_5$ modify the way different individual actions, once infected, affect future transmission. 

The epidemic model requires an initial seeding of infected individuals (controlled by $\theta_2$; see Table \ref{tab:params}). Once initialized, agent-based epidemic model accumulates infected individuals who may go on to infect others.  The result of a single run of this agent based model produces a  
cumulative count of infected individuals over time.  The result is stochastic since any encounter between an infected and a susceptible individual has a probability of producing an additional infected.  Each frame of Figure \ref{fig:sims} shows 100 replications of the model output for different parameter settings.

%
%

%
\section{Bayesian Model Formulation}
\label{sec:bayesmodel}

The application involves an ABM 
(i.e. a computer model, or simulator) from which only a
limited number of simulations may be carried out.  This application uses a standard setup in Bayesian computer model calibration where an
ensemble of ABM runs are carried out at input settings $\bm{\theta}^*_1,\ldots,\bm{\theta}^*_m$.
This ensemble is combined with observations $y$ of the actual epidemic, using a Bayesian statistical modeling formulation.  This formulation includes the ABM input parameters $\bm{\theta}$,
additional hyperparameters governing the statistical model formulation, an observation error term, and a systematic error term accounting for the difference between the model output and the epidemic observations.  
This formulation then produces updated (posterior) uncertainties for these various parameters and modeling terms from which we can produce predictions, 
with uncertainty, for future behavior of the epidemic.

One key component of this formulation is modeling the ABM output $\eta(\bm{\theta})$ at settings $\bm{\theta}$ that are not part of the initial ensemble.  For this, we use a Gaussian process (GP).
The current problem is complicated by the random results produced when running the ABM -- repeated runs of the ABM at the same input setting produce a distribution of epidemic curves (see Figure \ref{fig:sims}).
This stochastic output from the ABM makes many standard formulations for Bayesian computer model calibration inappropriate.  In this section we describe the details of the model output and how we can adapt this output so that it can be modeled within a general Bayesian computer model calibration formulation, while still appropriately capturing the relevant uncertainties in the problem.

\subsection{Agent-based model runs and observations}
\begin{figure}[!ht]
\centering
\includegraphics[width=0.7\textwidth]{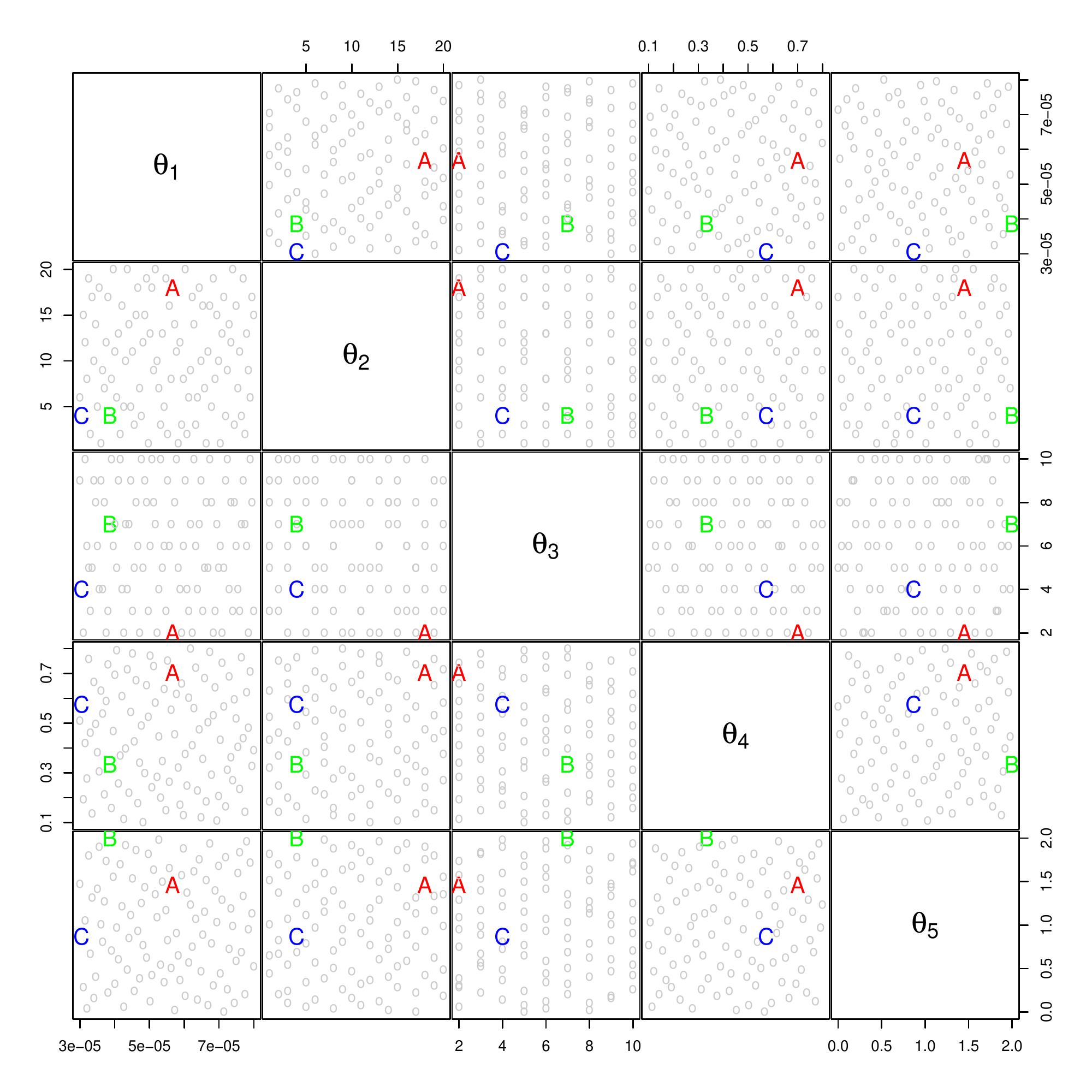}
\caption{2 dimensional projection of an $m=100$ point, OA-based LH design. The colored points are three chosen input settings which will be used later for holdout experiment.}
\label{fig:design}
\end{figure}

The ABM takes in a $p=5$-dimensional input parameter vector $\bm{\theta} = (\theta_1,\ldots,\theta_5)$ (Table \ref{tab:params}) in order to run.  We specify a collection of $m=100$ parameter settings $\bm{\theta}_1^*,\ldots,\bm{\theta}_m^*$, generated according to a space-filling, symmetric Latin hypercube design \citep{ye2000algorithmic}.  The 2-d margins of this $m \times p$ design are shown in Figure \ref{fig:design}. 

\begin{figure}[!bh]
\centering
\includegraphics[width=0.9\textwidth]{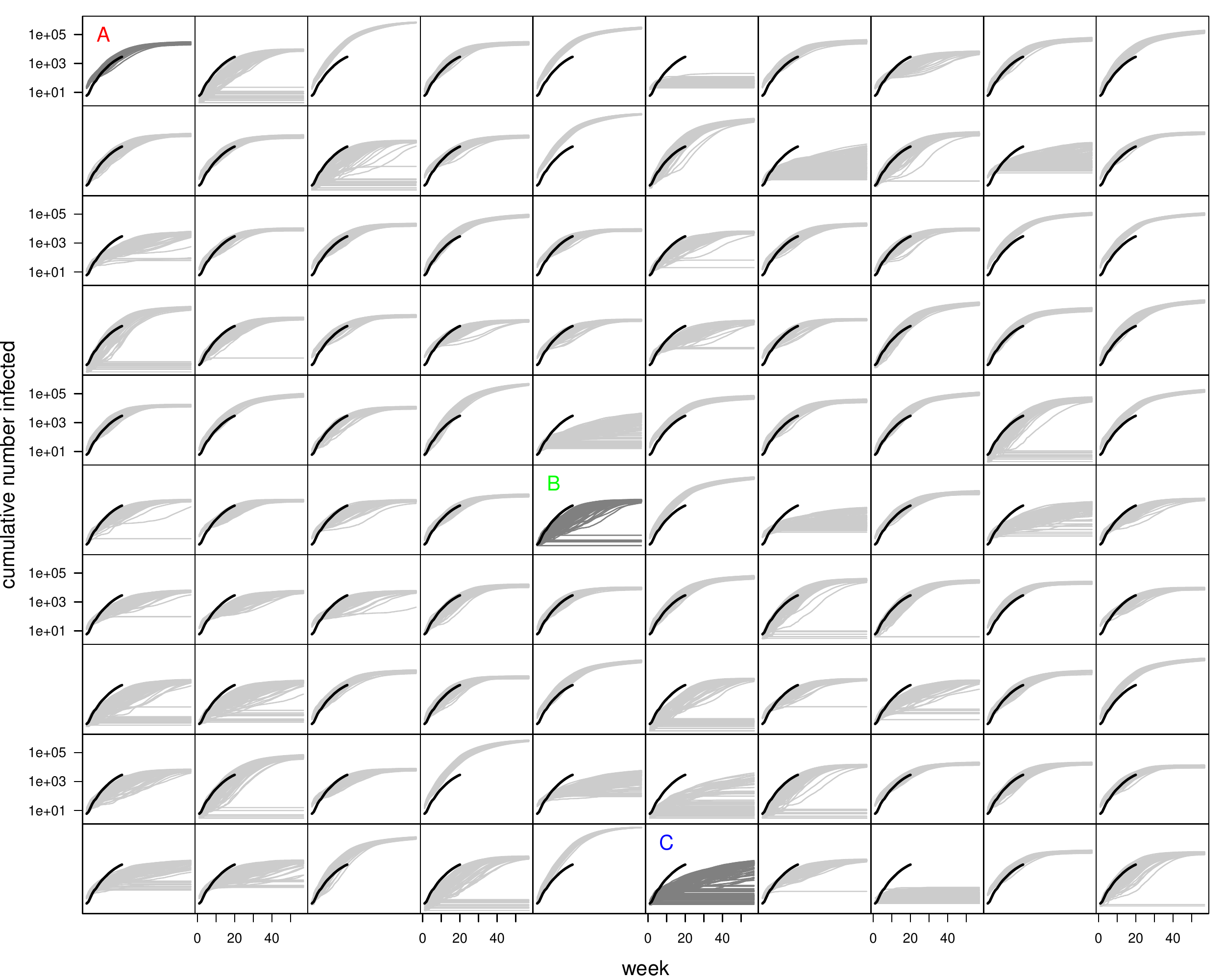}
\caption{The grey lines show the simulated trajectories of the cumulative number
of disease incidence for 57 weeks. Simulations in each box corresponds to 100 replicates of one input setting from the design in Fig \ref{fig:design}. Black lines are the 20 field observations. The three marked (by letters) boxes correspond to the design points which will be used later for the holdout experiment.}
\label{fig:sims}
\end{figure}

For each parameter setting, a collection of $r=100$ replicates were run, producing the output shown in Figure \ref{fig:sims}.  The log of the cumulative weekly number of infected individuals is taken as the model output, producing a 57-vector of logged counts. We were interested in studying the disease trajectory for a year, which justified the simulation of length 57.
In all, this gives an ensemble of 10,000 simulated epidemic trajectories, or curves.

We take the reported cases up to week 20 as the data
to be conditioned on for this calibration and prediction process illustration.  As with the simulations we take the log of the cumulative count by week.  These observed counts are given by the black line in each frame of Figure \ref{fig:sims}.

Note that for different parameter settings
the collection of replicate epidemics can show very different behavior.  
For some $\bm{\theta}_j^*$, the replicates are all tightly concentrated about a mean epidemic curve; in others, the replicates show a bimodal distribution -- some curves show the epidemic halting prior to week 20; others show an epidemic growing throughout the year.  Such variation in the distributions of the epidemic curve replicates as a function of $\bm{\theta}$ would make strategies that model the distributions as $N(\mu(\bm{\theta}),\sigma(\bm{\theta}))$ inappropriate.  
Hence we extend the quantile kriging approach of \citet{plumlee2014building} to emulate these epidemic curves, and embed this approach within the Bayesian calibration framework.  The basic concept is described in the next subsection.

\subsection{Defining quantiles and preprocessing}

From an examination of Figure \ref{fig:sims}, it is clear that for some input settings $\bm{\theta}$, only a small proportion of the replicates might be consistent with the observations.  For predicting plausible trajectories of the epidemic  after 20 weeks, it will be important to know the combination of replicates and parameter settings that are plausible given the data.

\begin{figure}[!th]
\centering
\includegraphics[width=1.0\textwidth]{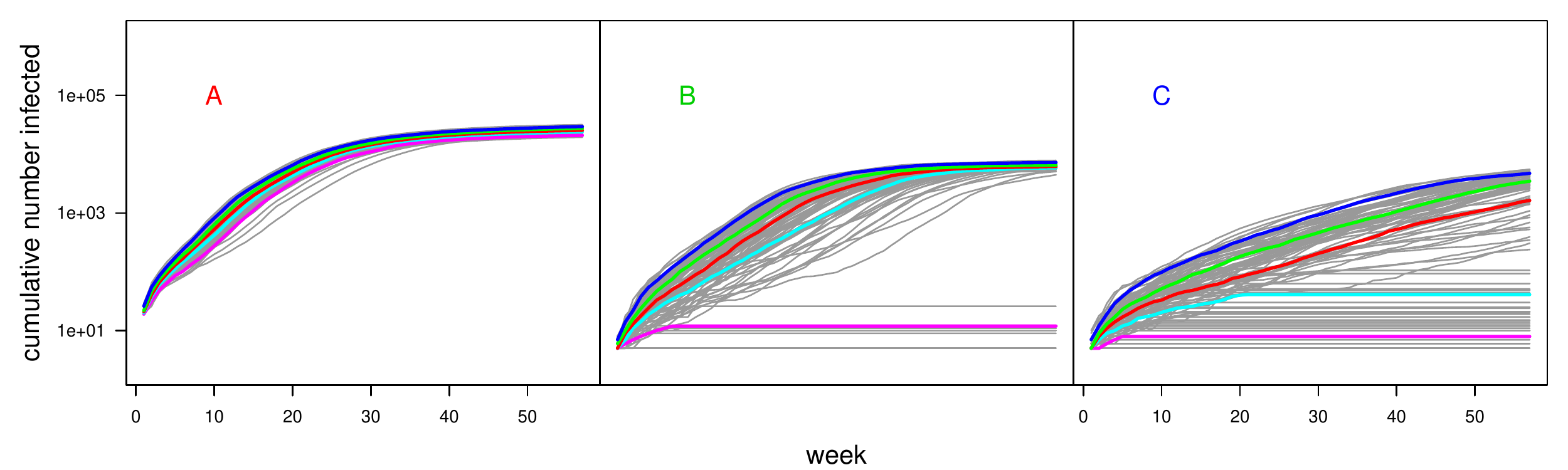}
\caption{The grey lines show the simulated trajectories of the cumulative number
of disease incidences for 57 weeks at three
different parameter settings $\bm{\theta}$.  The colored lines show the 5 estimated $.05,.275,.5,.725,.95$ quantiles.}
\label{fig:quantiles}
\end{figure}

This motivates us to find a way to index ABM-based epidemic curves by replicate, as well as by input parameter $\bm{\theta}$. In order to do this, we adapt the quantile kriging approach given in \citet{plumlee2014building}.  Figure \ref{fig:quantiles} shows the pointwise $\alpha = (.05,.275,.5,.725,.95)$-quantiles estimated from the 100 replicates for three different values of $\bm{\theta}$.
While these pointwise $\alpha$-quantiles (one for each of the 57 weeks in the simulation) do not correspond exactly any single ABM epidemic trajectory, they do produce curves very comparable to simulated epidemics.

Even though the replicates are i.i.d.~for a given $\bm{\theta}$, the quantile value $\alpha$ indexes continuous changes in the quantile process (see \citet{plumlee2014building} for necessary conditions and a proof).  By augmenting our formulation with the quantile value $\alpha$, we can consider the {\em effective} computational model that produces the epidemic count over 57 weeks
\[
  \eta(\bm{\theta},\alpha) = 
  (\eta_1(\bm{\theta},\alpha),\ldots,\eta_{57}(\bm{\theta},\alpha)),
  \mbox{ where }
\eta_t(\bm{\theta},\alpha) = 
  \inf \left\{x:P(\eta_t(\bm{\theta}) \leq x)
    \geq \alpha \right\}.
\]

We transform the initial ensemble of model runs consisting of $n_r=100$ replicates for each of $m=100$ parameter settings to a reduced ensemble of $n_\alpha=5$ quantiles for each of $m=100$ parameter settings.  This new ensemble can now be indexed by an effective parameter vector ($\bm{\theta}$,$\alpha$) of size $p+1$.

Hence we have the raw materials to carry out a calibration and prediction analysis in a number of different modeling frameworks
\citep{bayarri2007computer, 
drignei2008parameter,
higdon2008cmc, 
paulo2012calibration}.  
This includes 1) a $(m \cdot n_\alpha) \times (p+1)$ design of input parameter settings, and 2) a corresponding collection of model outputs (57-vector of weekly cumulative logged counts), one for each input setting, and 3) a collection of  observations giving the cumulative log counts for the first 20 weeks of the epidemic.  Our choice of using GPMSA was motivated by its use in the Ebola Challenge, as well as its inclusion of various nugget parameters to better account for the estimation error of the quantiles \citep{gattuq:2016}.  An outline of the model formulation encoded in GPMSA is given in the next subsection.

Finally, we note that our choice of $n_\alpha=5$ is based on exploring results using a variety of choices for $n_\alpha$.  A smaller value will reduce the computational effort of fitting our emulator and eventual formulation; a larger value means that the emulator need not interpolate over larger distances in the $\alpha$-component of the input space. Our explorations showed that the results do not change appreciably for $n_\alpha \geq 5$. Hence our choice came down to computational considerations.

\subsection{Bayesian model formulation}


The basic implementation in GPMSA \citep{gattuq:2016} models the observation vector $y$ as an additive combination of the computational model $\eta(\bm{\theta},\alpha)$ at the best setting for the parameters, a systematic model discrepancy term $\delta$, and observation error $\epsilon$:
\begin{equation}
\label{eq:model}
 y = \eta(\bm{\theta},\alpha) + \delta + \epsilon .
\end{equation}
When considering estimation, we can think of each of these terms being 20-vectors, spanning the range of the observations.  For prediction purposes, we can think of $\eta(\bm{\theta},\alpha)$, $\delta$ and $\epsilon$ as 57-vectors, predicting over the entire time period under consideration.  The resulting predictions are uncertain due to not knowing the parameters values $(\bm{\theta},\alpha)$, having a limited ensemble ABM runs so that $\eta(\bm{\theta},\alpha)$ must be estimated for settings not used in creating the ensemble, uncertain hyperparameters used in the statistical model, systematic error $\delta$ between the ABM and the actual epidemic's evolution, and error in the observations, modeled by $\epsilon$.  Priors for each of these terms are specified in GPMSA, with some modification allowed from the user.  This is briefly described next.

\begin{figure}[!ht]
\centering
\includegraphics[width=0.9\textwidth]{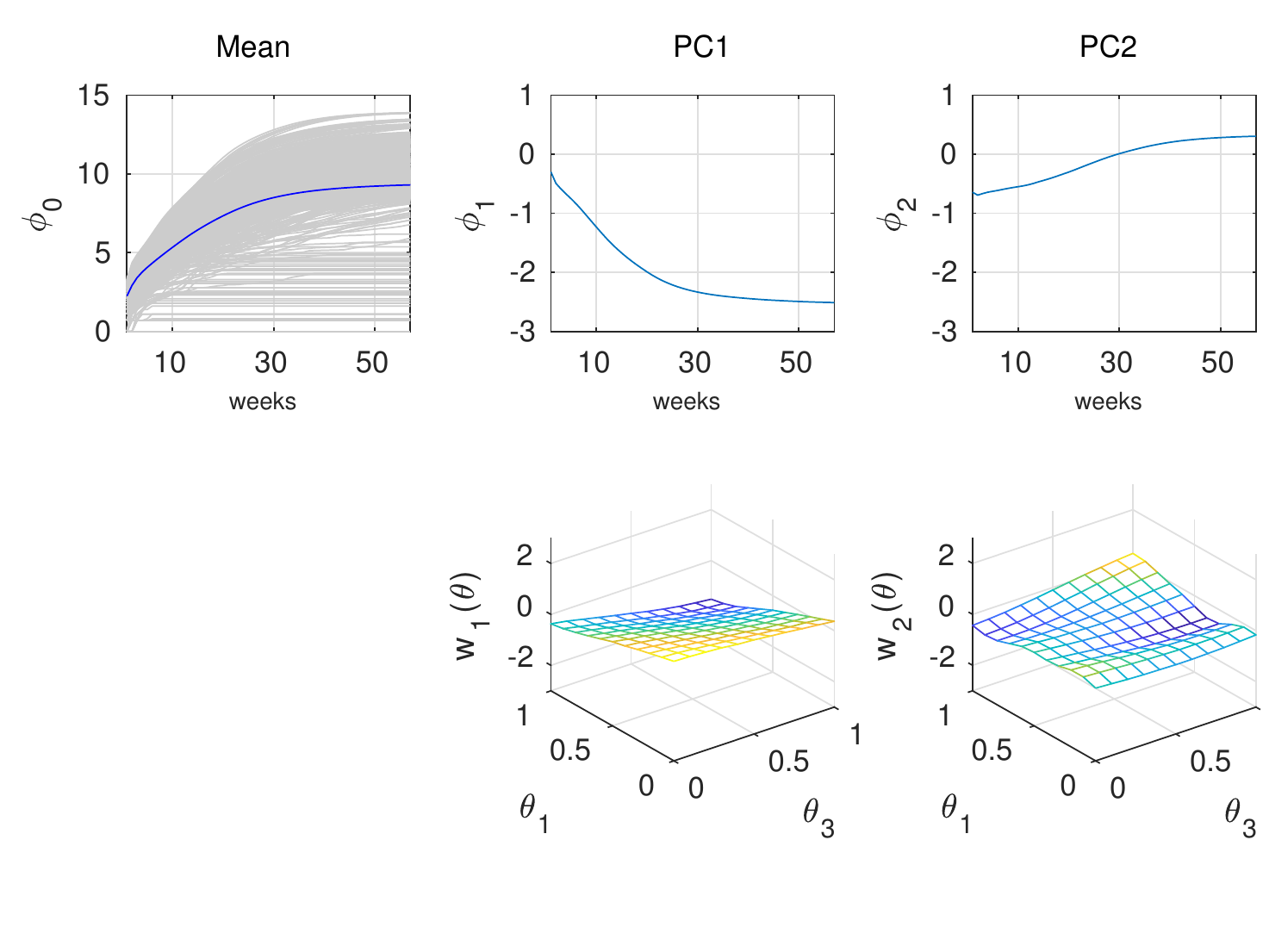}
\caption{Simulations (top left) and first two principal component bases (top right) obtained from the simulation output. The bottom row shows the posterior mean surface
for basis loadings $w_i(\bm{\theta}), i=1,2$ with respect to $\theta_1$ and $\theta_3$. Here the other fours parameters were held at their midpoints as $\theta_1$ and $\theta_3$ vary over the design range.}
\label{fig:wpred}
\end{figure}
A Gaussian process (GP) emulator is specified for the ABM output, so that inference regarding the model output at untried settings $\eta(\bm{\theta},\alpha)$ can be accounted for in the analysis.  
The GP prior allows one to incorporate uncertainty due to interpolation and prediction for this emulator.  In order to account for the multivariate nature of the model output, as well as the observations, a basis representation is specified for $\eta()$:
\[
 \eta(\bm{\theta}, \alpha) = \phi_0 + \sum_{k=1}^{p_\eta} \phi_k w_k(\bm{\theta}, \alpha) + \epsilon_{w0},
\]
where 
$p_\eta$ is the number of basis functions
used and $\epsilon_{w0}$ accounts for error due to the limited number of basis functions used.  Here we take $p_\eta=5$ for the analysis. The basis functions are eigenvectors (or empirical orthogonal functions \citep{eof:1999}) computed from the ensemble of epidemic curves. 
Figure \ref{fig:wpred} shows the mean of the ensemble mean $\phi_0$ along with the first two basis functions $\phi_1$ and $\phi_2$.  Also shown are posterior mean estimates for $w_1(\cdot)$ and 
$w_2(\cdot)$ in the $\theta_1$-$\theta_3$ plane.  The individual, univariate GPs $w_k(\bm{\theta})$ each have Gaussian covariance functions, with parameters controlling the marginal variance, the correlation length in each of the $p+1$ component directions, and a nugget term so that interpolation is not necessarily enforced, depending on the resulting posterior distribution. 
More details on the GP prior construction and its estimation can by found in the { appendix}.

The prior distribution for the input parameters and the quantile $(\bm{\theta},\alpha)$ is taken to be uniform over the hyper-rectangle given by the ranges in Table \ref{tab:params}, and a range of $[0,1]$ for $\alpha$.

The discrepancy term is modeled as a smoothing spline over time: 1 week $\leq$ time $\leq$ 57 weeks \citep{higdon2002space}.  The spline bases are normal kernels with a standard deviation of 18 weeks.  The marginal variance of the basis coefficients is estimated within GPMSA.  For this example, the resulting posterior for the variance is close to zero, so the impact of this term rather slight given the 20 weeks of observations (Figure \ref{fig:decomp}).  We also found that specifying kernel sds between 13 and 23 weeks did not appreciably change this result.

How to construct a likelihood function for such epidemic data is not straight-forward.  Here the data arrive as reported counts aggregated up from  surveillance records from the various counties in Liberia.  The challenging conditions under which the surveillance was carried out mean the reported counts could deviate substantially from the actual number of new cases in the population 
\citep{team2014ebola}.  
Also, counts early in the epidemic are likely to be low as general awareness has not yet pervaded health professionals, or the general population \citep{cori2017key,
owada2016epidemiological}.
The observation error is specified as a compromise between the L2 objective function used in 
\citet{venkatramanan2017using} 
and the Gaussian error term required by GPMSA.
Given the cumulative reported cases recorded in the $n_y=20$-vector $y$, we model the log likelihood as
\[
\ell(\bm{\theta},\alpha,\lambda_y;
  y,\eta(\cdot),\delta,\Sigma_y) = 
\frac{n_y}{2} \log \lambda_y -\frac{1}{2}
(y-\eta(\bm{\theta},\alpha)-\delta)^T
  \lambda_y \Sigma_y^{-1} (y-\eta(\bm{\theta},\alpha)-\delta) .
\]
The log-likelihood for the ABM parameters
and the precision scaling term $\lambda_y$ depends on the GP emulator $\eta(\cdot)$, the discrepancy vector $\delta$, and the $n_y \times n_y$ covariance matrix $\Sigma_y$.

The matrix $\Sigma_y$ is constructed so that the the weekly (non-cumulative) counts $c=(c_1,\ldots,c_{20})^T$ are 
treated as independent, with a sd of $\sigma_k=\max(5,0.2 \cdot c_k)$.  Thus the transformation from weekly counts $c$ to the log of the cumulative counts $y$ induces a transformation from $V = \mbox{Var}(c)$ to the covariance for the error term $\Sigma_y$ in (\ref{eq:model}). The variance scaling parameter $\lambda_y$ is also estimated in GPMSA.  

This likelihood term for the observations is augmented with a likelihood for the simulated epidemic curves (\ref{eq:like1}).  The Bayesian formulation is completed by the prior specification outlined in the appendix. The prior parameters include the ABM input parameters as well as a host of parameters controlling various GP and error terms in this specification.


\section{Results}

\begin{figure}[!ht]
\centering
\includegraphics[width=0.8\textwidth]{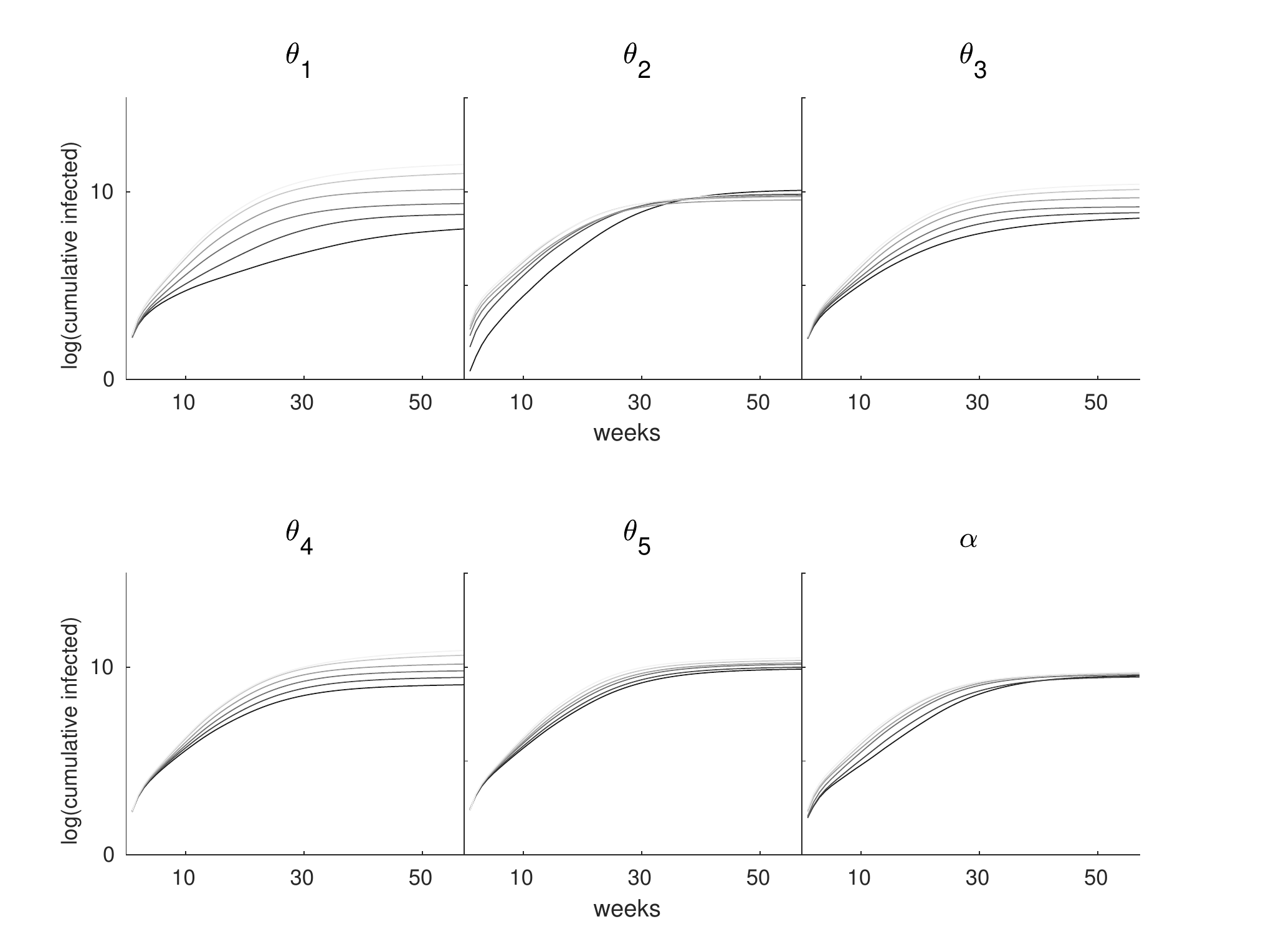}
\caption{Posterior mean simulator predictions (log counts as a function of time) varying one input across its prior range, holding others at their posterior mean values.}
\label{fig:thetasens_new}
\end{figure}

Figure \ref{fig:thetasens_new} shows posterior means for the
simulator response $\eta(\cdot)$ where each of the six inputs
were varied over their prior range while the
other five inputs were held at their posterior mean values.
The posterior mean response surfaces convey an idea of how the different
parameters affect the multivariate simulation output near the posterior mean.
Other marginal functionals of the simulation response can also
be calculated such as sensitivity indicies or estimates of the
Sobol decomposition
\citep{sack:welc:mitc:wynn:1989,oakl:ohag:2004}.
%

While Figure \ref{fig:thetasens_new} shows main effects of the parameters on model output, it is clear from Figure \ref{fig:sims} that the quantile's effect on the model output clearly depends on the size of the epidemic which is determined by other model inputs.  For instance, at the input setting marked by ``A'' the quantile has very little impact; for settings ``B'' and ``C'', the epidemic curves change substantially with quantile.


\begin{figure}[th!]
\centering
\includegraphics[width=1\textwidth]{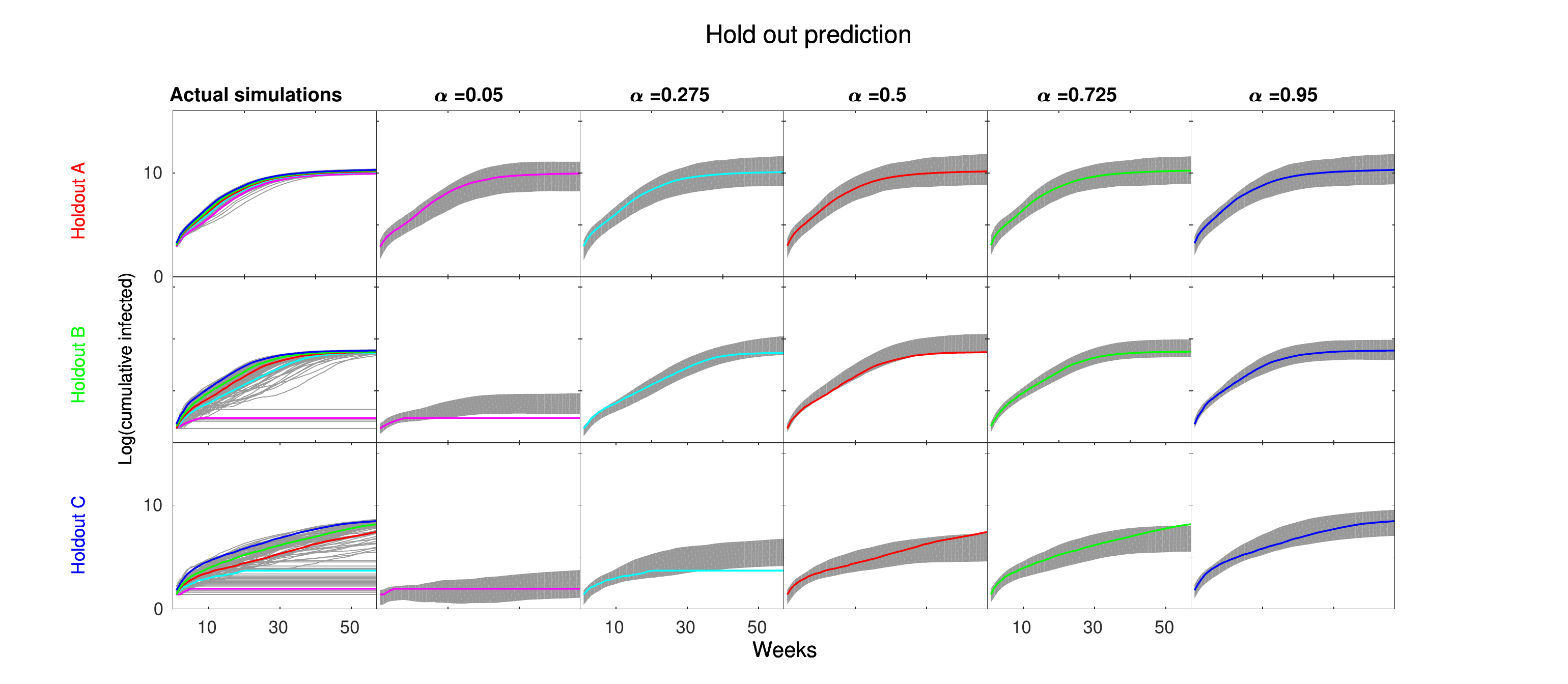}
\caption{The first column from the left shows the actual simulations and their empirical quantiles at the three design inputs denoted by their colors from the design in Fig \ref{fig:design}. Next 5 columns show the pointwise 90\% credible intervals of the posterior emulator prediction at those model inputs.  These are genuine holdout predictions since these AMB runs were not used to train the GP based emulator. Each of these five columns compares the empirical quantiles, denoted by different colors, from the actual simulations with the posterior predictions.}
\label{fig:holdout}
\end{figure}
We can assess the accuracy of this quantile based emulator by predicting the simulated trajectories of log of cumulative disease incidence for three holdout design points shown in Figures \ref{fig:design} and \ref{fig:sims}. The resulting simulated and predicted trajectories are shown in Figure \ref{fig:holdout}. Each row of the figure corresponds to one of the design points -- A, B, and C. The first column shows the 100 replicates, along with the estimated quantiles. The remaining 5 columns are the plots of pointwise 90\% credible intervals for each of the five quantiles, along with the quantiles estimated from the ABM replicates.  

\begin{figure}[!ht]
\centering
\includegraphics[width=0.85\textwidth]{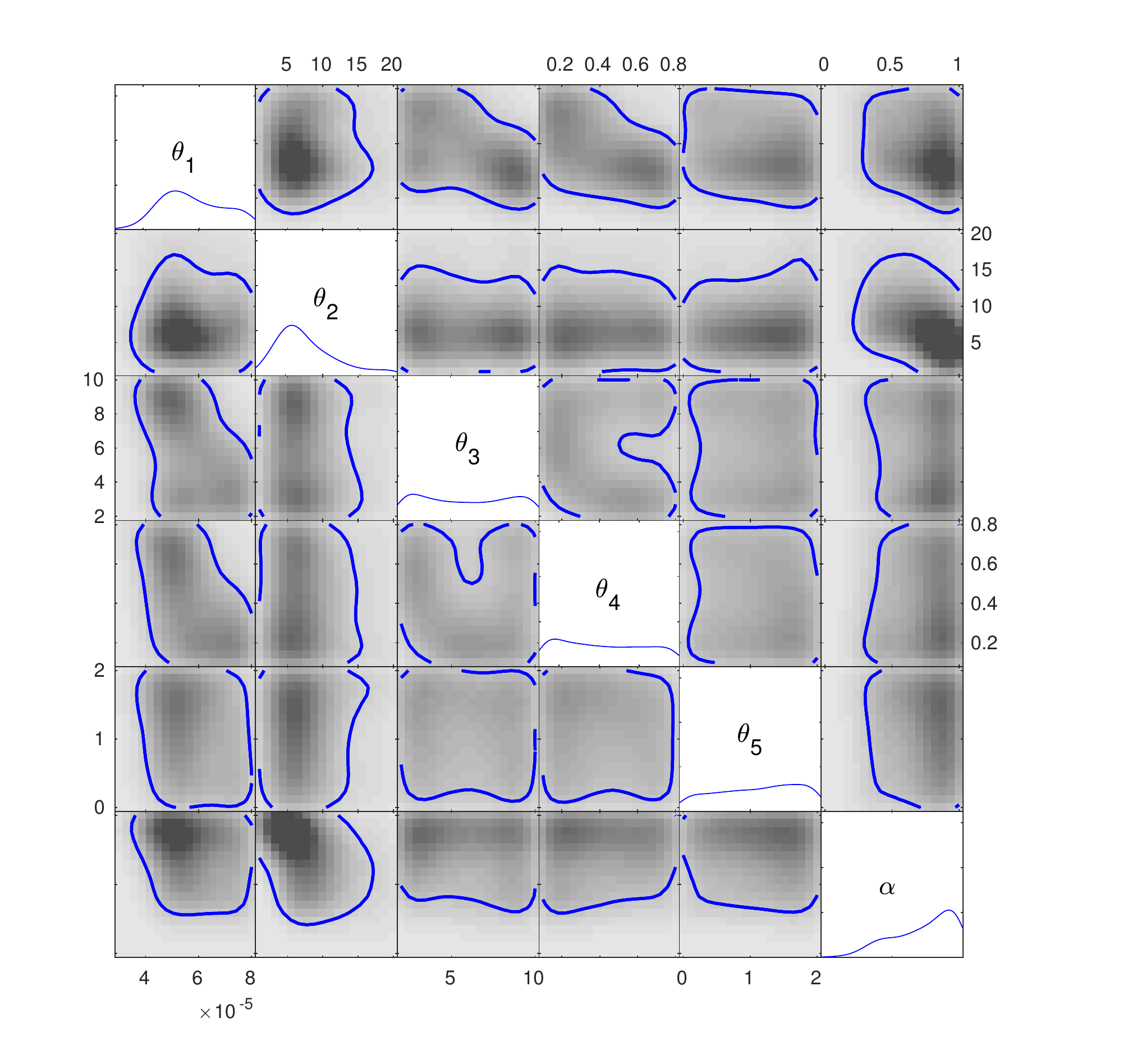}
\caption{Estimated posterior distribution of the parameters $(\theta_1, \cdots, \theta_5,
\alpha)$. The diagonal shows the estimated marginal posterior pdf for each parameter; the off-diagonal images give estimates of bivariate marginals; to contour lines show estimated 90\% hpd regions.}
\label{fig:thetapost}
\end{figure}
The posterior for the input parameters, along with the quantile, $(\bm{\theta},\alpha)$, is depicted in Figure \ref{fig:thetapost}.  This shows the 1-d and 2-d margins of this 6-d posterior distribution.  The parameters controlling the transmissibility $\theta_1$, the initial number of infected individuals $\theta_2$, and the quantile $\alpha$ are constrained by the observations up to week 20.

\begin{figure}[ht!]
\centering
\includegraphics[width=0.99\textwidth]{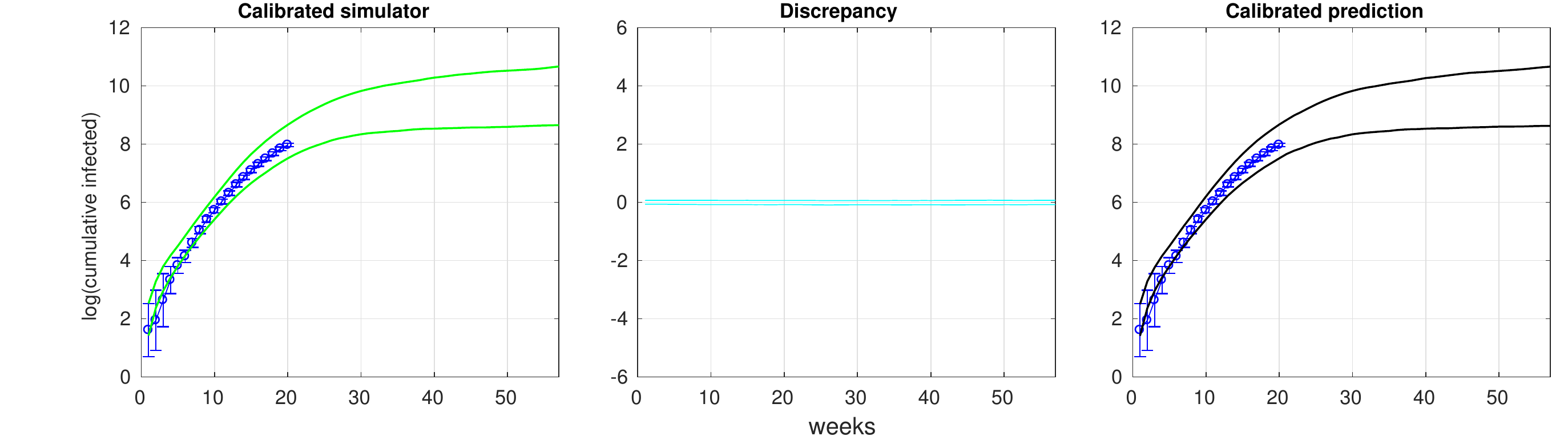}
\caption{Decomposition of the posterior prediction for the epidemic curve.  Left: posterior 90\% intervals for the calibrated emulator $\eta(\bm{\theta})$.  The intervals are due to uncertainty regarding both $\bm{\theta}$ and the GP emulator of ABM.  Middle: posterior 90\% intervals for $\delta$.  Right: posterior 90\% intervals for the actual epidemic curve: $\eta(\bm{\theta}) + \delta$.  The blue dots show the observed counts, along with 1 sd bars given by the square root of diag$(\Sigma_y)$.}
\label{fig:decomp}
\end{figure}
The posterior pointwise 90\% intervals for the calibrated ABM $\eta(\bm{\theta})$ and discrepancy $\delta$ are given in the first two frames of Figure \ref{fig:decomp}.  The final frame shows the 90\% intervals for actual epidemic $\eta(\bm{\theta}) + \delta$. 
Even though the posterior uncertainty for $\delta$ is comparatively small here, the terms $\eta(\bm{\theta})$ and $\delta$ are negatively correlated.  This means the posterior variance for the actual epidemic is less than the sum of the posterior variance of these two components.

\begin{figure}[ht]
\centering
\includegraphics[width=0.99\textwidth]{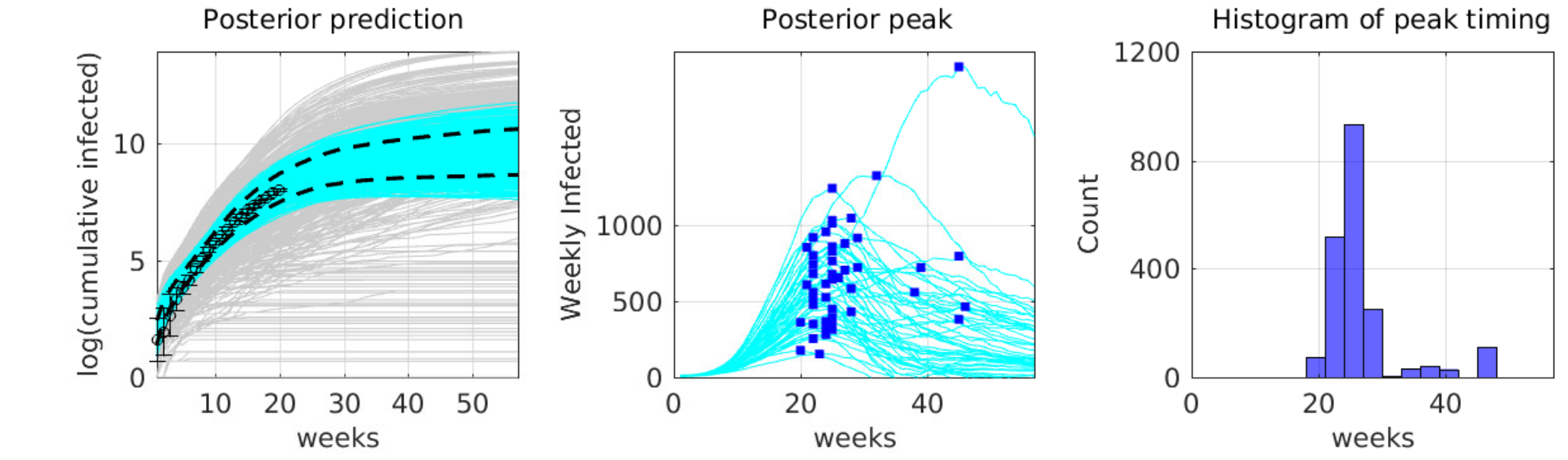}
\caption{Left: Posterior (cyan) and prior (gray) realizations of the  epidemic curves. The dashed line gives the pointwise 90\% credible intervals for the epidemic curves. Middle: 40 posterior realizations of the weekly counts over time.  The blue dots show the peak weekly count and corresponding time. Right: histogram of posterior peak time realizations.}
\label{fig:pred}
\end{figure}

While the constriction in moving from prior
to posterior may not look substantial from examining Figure \ref{fig:thetapost}, the resulting reduction in uncertainty is apparent when examining the posterior epidemic curves (Figure \ref{fig:pred}, left frame).  The posterior distribution of the epidemic curves accounts for uncertainty in the parameters, model discrepancy, uncertainty in the emulation of the ABM, and observation error. 
From these posterior realizations, one can also produce posterior uncertainties for a number of other quantities, such as the timing and count of the peak week of the epidemic.  The blue dots in the middle frame of Figure \ref{fig:pred} show posterior realizations of the peak (time,infected) points; the cyan curve is obtained by exponentiating and differencing the cumulative curve in the left frame of figure.  The analysis shows that the peak of the epidemic is most likely to be between weeks 20 and 30, but the peak time plausibly could be as late as 48 weeks (right frame of Figure \ref{fig:pred}).

\subsection{Predictions for the Ebola Challenge}

The Ebola Challenge required predictions and a description of their uncertainty for some of the key quantities 
\citep{tabataba2017framework}
weeks 13, 26, 35, and 42.  We focus on three of these quantities: 
\begin{itemize}
\item
Peak timing -- the week in which the epidemic increased by the highest number of cases;
\item
Peak weak cases: the number of new infected individuals incurred on the peak timing week;
\item
Total epidemic size -- the total number infected after the 57-week time period.
\end{itemize}
\begin{figure}[ht]
\centering
\includegraphics[width=0.99\textwidth]{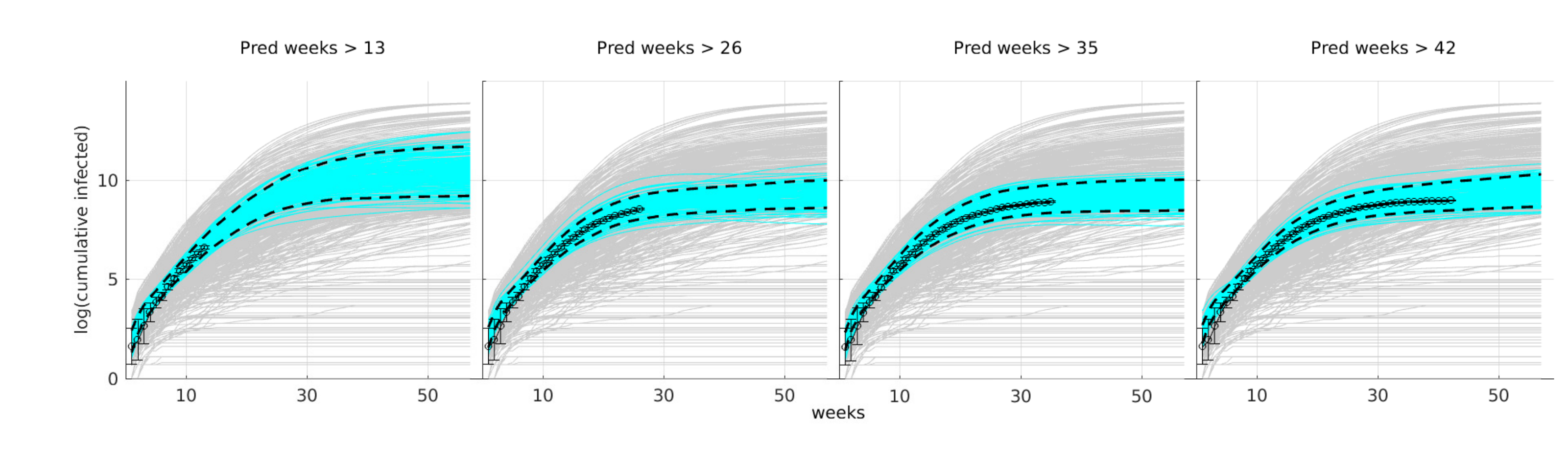}
\caption{Posterior draws (cyan lines) and pointwise 90\% credible intervals (dashed lines) for the actual epidemic curves ($\eta(\theta) + \delta$) for each of the ebola challenge time periods.  The gray lines show epidemic curves produced from the initial ABM ensemble.}
\label{fig:postPredChallenge}
\end{figure}

\begin{figure}[ht]
\centering
\fbox{\includegraphics[width=0.9\textwidth]{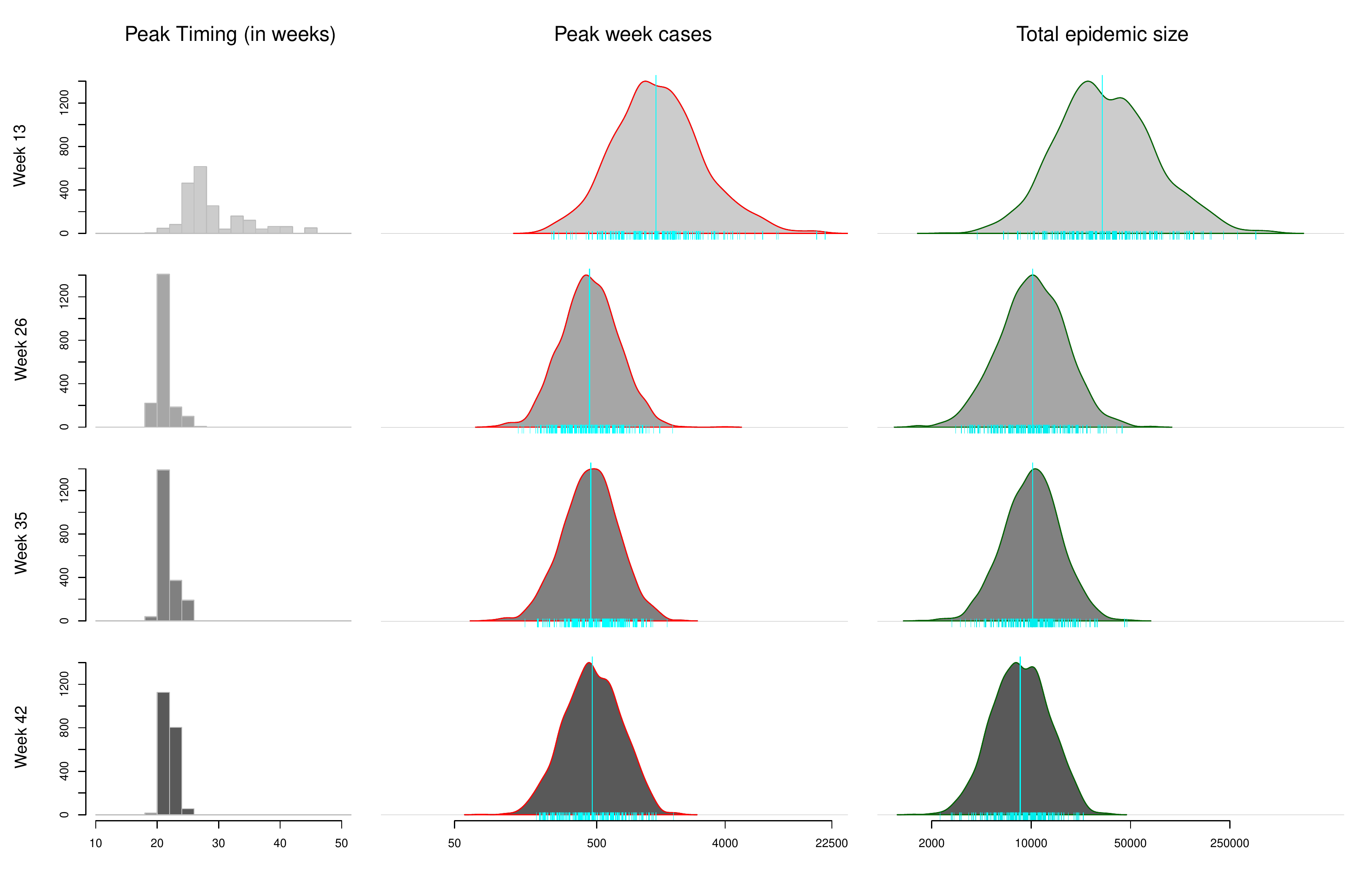}}
\caption{Posterior distributions for three functions of the actual epidemic curve -- peak timing (left), peak weakly cases (middle), and total epidemic size (right).
Each row shows the posterior density estimates conditional on epidemic data
up to 13, 26, 35 and 42 weeks, moving from top to bottom.}
\label{fig:postChSum}
\end{figure}

Figure \ref{fig:postPredChallenge} shows posterior realizations for the actual epidemic curve conditional on the data used at each of the timepoints.  From these, posterior draws for each of the quantities of interest above can be produced.  The marginal posterior distributions are depicted in Figure \ref{fig:postChSum}.

Note that even after 42 weeks worth of observations, the uncertainty regarding the actual epidemic curve is still substantial. One reason for this is that the epidemic is mostly over by week 30, so that observations after that time don't add information to the posterior.  
This uncertainty is
also influenced by the estimated covariance matrix $\lambda_y^{-1}\Sigma_y$ that links the actual epidemic to the observations $y$. 

Given this, it's not surprising that the posterior distributions for peak timing, peak week cases, and total size are fairly stable by the second time period in the ebola challenge (bottom three rows of Figure \ref{fig:postChSum}).  A closer inspection of these posterior distributions shows that additional observations beyond week 26 serve to eliminate extreme values for timing and cases that are still plausible in the posterior produced at week 26.

We also note that the role of the discrepancy term $\delta$ becomes slightly more prominent for analyses that involve more weeks of observations.  
This is due to the fact that none of the ABM epidemic curves are fully compatible with the actual one.  So while the posterior prediction for $\eta(\bm{\theta}) + \delta$ takes $\delta$ to be nearly 0 at the early time periods, later time periods require more contribution from $\delta$. This affects the posterior for $\bm{\theta}$ as well.  When there is a mismatch between model and reality, this trade-off between $\eta(\bm{\theta})$ and $\delta$ is well known, and limits the ability of the data alone to reduce uncertainty regarding $\bm{\theta}$
\citep{kenn:ohag:2001,higd:kenn:cave:2004,bayarri2007cmv,brynjarsdottir2014learning,tuo2016theoretical}.

\section{Discussion}

This paper shows how the quantile kriging approach can be extended to handle multivariate model output, and how it can be embedded into a Bayesian computer model calibration framework that accounts for uncertainty in calibration parameters, parameters governing the GP emulator, systematic errors between the model and reality, and the stochastic nature of the ABM.  For a given ABM input setting $\bm{\theta}$, a GP effectively gives a non-parametric model for the quantiles of the epidemic curve replicates.
The preprocessing described here shows how one can use existing modeling frameworks and software to handle this problem.

This strategy allows rather quick estimation for an application that entails a substantial amount of simulation output -- 100 5-d input settings $\times$ 100 replicates $\times$ 57-week simulation output.  Because of this, we were able to carry out the very demanding analyses and reporting that was required for the Ebola Challenge.

This solution is different than producing emulators for the mean and covariance for the replicates as a function of $\bm{\theta}$.  While emulating the mean is straightforward, how to emulate a covariance matrix as a function of the 5-d input space is not.  In addition, such an approach would likely require a normality assumption for the replicate variation -- it is clear from some of the frames in Figure \ref{fig:sims} that a normal approximation is insufficient at some input settings $\bm{\theta}$.  Hence some classification approach (e.g. \citet{crevillen2017gaussian}) or mixture of Gaussians would be necessary.  

The applications described here reveal some shortcomings as well.  While the GP used here nicely accounts for dependencies between quantiles, it does not enforce monotonicity: 
$\eta(\bm{\theta},\alpha_1) \leq \eta(\bm{\theta},\alpha_2)$ if $\alpha_1 \leq \alpha_2$.
Approaches for ensuring monotonicity are available \citet{lin2014bayesian,wang2016estimating}, but their integration into general Bayesian model calibration formulations remains a research topic.

Also, for replicated, multivariate model output, how one defines and estimates the quantile is still a bit of an art.  For this application, using the pointwise quantile over each of the 57 weeks was fine since the resulting quantile trajectories were quite compatible with the actual ABM epidemic curves.  In other cases, one might need more care in ordering the multivariate model output appropriately so that a coherent mapping of output to quantiles is made, and that the resulting mapping leads to something well modeled by a GP over the $(\bm{\theta},\alpha)$ input space. 
 
We end by noting that ABMs are becoming increasingly common for modeling social systems at the individual level, incorporating structure, constraints and behaviors, while 
producing very detailed -- but stochastic -- output.  This paper gives one approach for producing inferences with the aid of such models; others include 
\citet{hooten2010statistical,wikle2016hierarchical,andrianakis2017efficient,kattwinkel2017bayesian}.
The allure of such models is their direct connection to reality (e.g.  with an ABM one can directly compute the impact of closing schools on the evolution of the epidemic). The challenge is to appropriately quantify the uncertainty in how such models differ from reality.  

 \section*{Acknowledgements}
This work was supported in part by a number of granting offices.  These include:
IARPA Embers grant 432958; DTRA CNIMS program, grant 450117; the NSF NetSE program, grant 478403; and the NSF DIBBS program, grant 417549; and
the U.S. Department of Energy Office of Science, Office of Advanced Scientific Computing Research, Scientific Discovery through Advanced Computing (SciDAC) program.


%

\section*{Appendix: Quantile Kriging}

By the term {\em kriging}, we mean simple interpolation of a function at a given input, where the functional value
is unknown. {\em Quantile kriging} is a technique of modeling the distribution of model output via modeling its quantile
as a function of the input. Instead of modeling a stochastic simulator by fitting a deterministic mean function and a zero mean gaussian error in an additive set up, here we model the $\alpha$ quantile $Q_\alpha$ of the simulation output $Y(x)$
at a give input $x$ as a function of $x$. Given any $\alpha$ quantile exists for some $x \in X$, one can model
$Q_\alpha(x_0)$ for an unobserved point $x_0 \in X$ using simple gaussian process regression.  The continuity assumption
needed for a GP regression is well satisfied for the quantile functions. Once we fit $Q_\alpha$ for different $\alpha$,
the cumulative distribution of simulation output can easily be written down using the quantile estimates \citep{plumlee2014building}.

\subsection*{GPMSA Formulation}

The ABM is run at $m=100$ input settings varying over predefined ranges of the $p=5$ dimensional input variables given by Table \ref{tab:params}. The $(m \cdot n_\alpha) \times (p + 1)$ dimensional design matrix including the column of $n_\alpha=5$ quantiles is given as:
\begin{equation}
\label{eq:design}
\begin{pmatrix}
\theta_{11}^* & \cdots & \theta_{1p}^* & \alpha_1
\cr \vdots        & \ddots & \vdots & \vdots \cr
\theta_{11}^* & \cdots & \theta_{1p}^* & \alpha_{n_\alpha}
\cr \vdots        & \ddots & \vdots & \vdots \cr
\theta_{m1}^* & \cdots & \theta_{mp}^* & \alpha_1
\cr \vdots        & \ddots & \vdots & \vdots \cr
\theta_{m1}^* & \cdots & \theta_{mp}^* & \alpha_{n_\alpha}
\end{pmatrix}.
\end{equation}
Quantile estimates from $r=100$ trajectories of cumulative number of infected persons, at each of the $m$ input settings, as seen in Figure \ref{fig:sims}, is considered as the output from ABM.  For this appendix, we subsume the quantile value $\alpha$ into the parameter vector $\bm{\theta}$
and take $m$ to denote the number of input settings. Thus 
the design (\ref{eq:design}) is taken to be $m \times p$.  Also, the dimension $p_\theta$ of the input vector is now larger by one.
 
The observation vector $y$ is modeled as the sum of a computational model $\eta(\bm{\theta})$, a systematic model discrepancy term $\delta$, and observational error $\epsilon$ as described in equation \ref{eq:model}. The computational model $\eta(\bm{\theta})$ is represented as linear combination of $p_\eta$ basis functions, and an overall mean.
\[ \eta(\bm{\theta}) = \phi_0 + \sum_{i=1}^{p_\eta} \phi_i w_i(\bm{\theta})
+\epsilon_{w0}, \]
where $\theta_6$ represents the quantile parameter $\alpha$.  The vector 
$\epsilon_{w0}$ is given a 
$N(0,\lambda_{w0}^{-1} I)$ prior distribution.
Each basis weight $w_i(\bm{\theta})$, $i=1,\dots,p_\eta$, is then
modeled as a mean 0 GP
\begin{equation*}
  w_i(\bm{\theta}) \sim
\mbox{GP}(0,\lambda^{-1}_{wi} R(\bm{\theta}, \bm{\theta}';\rho_{wi})
),
\end{equation*}
where $\lambda_{wi}$ is the marginal precision of the process and
the correlation function is given by
\begin{equation*}
\label{eq:wcor}
R(\bm{\theta},\bm{\theta}';\rho_{wi}) =
  \prod_{k=1}^{p_\theta} \rho_{wik}^{4(\theta_{k} - \theta'_{k})^2}.
\end{equation*}
The parameter $\rho_{wik}$ controls
the spatial range for the $k$th input dimension of the process
$w_i(\bm{\theta})$. 
The scaling of each $\phi_i$ is chosen so that the marginal variance of $w_i(\cdot)$ should be close to 1.  Hence the priors for the $\lambda_{wi}$s have mass centered at 1. 
Under this parameterization, $\rho_{wik}$ gives the
correlation between $w_i(\bm{\theta})$ and $w_i(\bm{\theta}')$ when the input
conditions $\bm{\theta}$ and $\bm{\theta}'$ are identical, except for a
difference of half the prior range of the $k$th component. More details about the covariance function can be found in \citep{gattuq:2016}.

If we restrict to the $m$ input settings used for the ensemble we can define,
\[
 w_i = (w_i(\bm{\theta}^*_{1}),\ldots,
        w_i(\bm{\theta}^*_{m}))',\;\;
 i=1,\ldots,p_\eta.
\]
Then, $w=(w_1',\ldots,w_{p_\eta}')'$ then has prior distribution
\begin{equation}
\label{eq:wprior1}
  \begin{pmatrix} w_1 \cr \vdots \cr w_{p_\eta} \end{pmatrix}
  \sim
  N\left( \begin{pmatrix} 0 \cr \vdots \cr 0 \end{pmatrix},
  \Sigma_w = \begin{pmatrix}
     \lambda^{-1}_{w1}R(\bm{\theta}^*;\rho_{w1})
                                      & 0 & 0 \cr
     0 & \ddots & 0 \cr
     0 & 0 & \lambda^{-1}_{wp_\eta}R(\bm{\theta}^*;\rho_{wp_\eta})
  \end{pmatrix}
  \right),
\end{equation}
where $R(\bm{\theta}^*;\rho_{wi})$ is obtained by applying the Gaussian covariance function on the design matrix in (\ref{eq:design}). 
The ABM output $\eta(\bm{\theta}_j^*)$ is projected (via dot product) onto each basis vector $\phi_i$, giving transformed output 
$w_i^*=(w^*_i(\bm{\theta}_1^*),\ldots,w^*_i(\bm{\theta}_m^*))$.
These transformed simulations are then modeled as independent normal perturbations from the $w_i$ vectors: 
\begin{equation}
\label{eq:like1}
w^*_i \sim N(w_i,\lambda_{w\epsilon i} I_m), 
i=1,\ldots,p_\eta.
\end{equation}
In standard settings, the ``nugget'' precisions $\lambda_{w\epsilon i}$ are typically quite large, so that the posterior produces $w_i(\cdot)$s that essentially interpolate the projected simulations $w^*_i$.  Since the quantiles are estimated from the 100 replicates, the posterior values for $\lambda_{w\epsilon i}$ don't enforce exact interpolation.

We specify independent, diffuse gamma priors for $\lambda_{w\epsilon i}$ and $\lambda_{w0}$, independent $\Gamma(5,5)$ priors for each $\lambda_{wi}$, and
independent beta$(a_{\rho_w},b_{\rho_w})$ priors for the $\rho_{wik}$'s.
\begin{eqnarray*}
\pi(\lambda_{w0}) & \propto & \lambda_{w}^{a_{w0}-1} e^{-b_{w0} \lambda_{w0}}, \\
\pi(\lambda_{w\epsilon i}) & \propto & \lambda_{w \epsilon i}^{a_{w \epsilon}-1} e^{-b_{w \epsilon} \lambda_{w \epsilon i}},
   \;\; i=1,\ldots,p_\eta, \\
   \pi(\lambda_{wi}) & \propto & \lambda_{wi}^{a_w-1} e^{-b_w \lambda_{wi}},
   \;\; i=1,\ldots,p_\eta, \\   
\pi(\rho_{wik}) & \propto & \rho_{wik}^{a_{\rho_w}-1} (1-\rho_{wik})^{b_{\rho_w}-1},
   \;\; i=1,\ldots,p_\eta, \; k=1,\ldots,p_\theta.
\end{eqnarray*}

Like the emulator, the discrepancy term $\delta$ is modeled using a basis representation over time: 1 week $\le$ time $\le$ 57 weeks.
However, the discrepancy basis vectors are 1-d normal kernels with an sd of 18 weeks; the kernels are spaced 12 weeks apart. 
\begin{equation}
\label{eq:discrep}
  \delta = \sum_{k=1}^{p_\delta} d_k v_k,
\end{equation}
where $p_\delta=7$ and each $v_k$ has a zero mean normal prior with precision $\lambda_\delta$. Combining this with the simulator model, the data log likelihood is then re-written as:
\begin{eqnarray}
\label{eq:like2}
\lefteqn{\,\,\,\,\,\,\,\, \ell(\theta,\alpha,\lambda_y,w(\cdot),v,\lambda_{w},\lambda_\delta,\rho_w,;
  y,\phi,d,\Sigma_y)  =   \frac{n_y}{2} \log \lambda_y} \\ 
 \nonumber 
 &&
 -  \frac{1}{2}
\Big(y-\phi_0 - \sum_{k=1}^{p_\eta} \phi_k w_k(\theta)-\sum_{k=1}^{p_\delta} d_k v_k \Big)^T  
(\lambda_y \Sigma_y^{-1} + \lambda_{w0} I)
 \Big(y-\phi_0 - \sum_{k=1}^{p_\eta} \phi_k w_k(\theta)-\sum_{k=1}^{p_\delta} d_k v_k \Big). \nonumber
\end{eqnarray}
The
 full specification of the likelihood follows from (\ref{eq:like1}) and  (\ref{eq:like2}).  
This, along with the prior specification outlined above gives the posterior distribution for the unknown parameters.  This distribution is sampled via MCMC in GPMSA, which can also produce draws from the posterior predictive distribution for the emulator and the epidemic curves
\citep{higdon2008cmc,gattuq:2016}.
\renewcommand{\baselinestretch}{1.0}
\footnotesize

\bibliography{dave}

\begin{thebibliography}{38}
\newcommand{\enquote}[1]{``#1''}
\expandafter\ifx\csname natexlab\endcsname\relax\def\natexlab#1{#1}\fi

\bibitem[\protect\citename{Andrianakis et~al., }2017]{andrianakis2017efficient}
Andrianakis, I., McCreesh, N., Vernon, I., McKinley, T., Oakley, J., Nsubuga,
  R., Goldstein, M., and White, R. (2017).
\newblock \enquote{Efficient history matching of a high dimensional individual
  based HIV transmission model.}
\newblock {\em ASA SIAM Journal on Uncertainty Quanitification\/}.

\bibitem[\protect\citename{Andrianakis et~al., }2015]{andrianakis2015bayesian}
Andrianakis, I., Vernon, I.~R., McCreesh, N., McKinley, T.~J., Oakley, J.~E.,
  Nsubuga, R.~N., Goldstein, M., and White, R.~G. (2015).
\newblock \enquote{Bayesian history matching of complex infectious disease
  models using emulation: a tutorial and a case study on HIV in Uganda.}
\newblock {\em PLOS Comput Biol\/}, 11, 1, e1003968.

\bibitem[\protect\citename{Bayarri et~al.,
  }2007{\natexlab{a}}]{bayarri2007computer}
Bayarri, M., Berger, J., Cafeo, J., Garcia-Donato, G., Liu, F., Palomo, J.,
  Parthasarathy, R., Paulo, R., Sacks, J., and Walsh, D. (2007{\natexlab{a}}).
\newblock \enquote{Computer model validation with functional output.}
\newblock {\em The Annals of Statistics\/},  1874--1906.

\bibitem[\protect\citename{Bayarri et~al., }2007{\natexlab{b}}]{bayarri2007cmv}
Bayarri, M., Berger, J., Cafeo, J., Garcia-Donato, G., Liu, F., Sacks, J., and
  Walsh, D. (2007{\natexlab{b}}).
\newblock \enquote{{Computer model validation with functional output}.}
\newblock {\em Annals of Statistics\/}, 35, 5, 1874.

\bibitem[\protect\citename{Brynjarsd{\'o}ttir and OʼHagan,
  }2014]{brynjarsdottir2014learning}
Brynjarsd{\'o}ttir, J. and OʼHagan, A. (2014).
\newblock \enquote{Learning about physical parameters: The importance of model
  discrepancy.}
\newblock {\em Inverse Problems\/}, 30, 11, 114007.

\bibitem[\protect\citename{Cori et~al., }2017]{cori2017key}
Cori, A., Donnelly, C.~A., Dorigatti, I., Ferguson, N.~M., Fraser, C., Garske,
  T., Jombart, T., Nedjati-Gilani, G., Nouvellet, P., Riley, S., et~al. (2017).
\newblock \enquote{Key data for outbreak evaluation: building on the Ebola
  experience.}
\newblock {\em Phil. Trans. R. Soc. B\/}, 372, 1721, 20160371.

\bibitem[\protect\citename{Cornuet et~al., }2008]{cornuet2008inferring}
Cornuet, J.-M., Santos, F., Beaumont, M.~A., Robert, C.~P., Marin, J.-M.,
  Balding, D.~J., Guillemaud, T., and Estoup, A. (2008).
\newblock \enquote{Inferring population history with DIY ABC: a user-friendly
  approach to approximate Bayesian computation.}
\newblock {\em Bioinformatics\/}, 24, 23, 2713--2719.

\bibitem[\protect\citename{Crevill{\'e}n-Garc{\'\i}a et~al.,
  }2017]{crevillen2017gaussian}
Crevill{\'e}n-Garc{\'\i}a, D., Wilkinson, R., Shah, A., and Power, H. (2017).
\newblock \enquote{Gaussian process modelling for uncertainty quantification in
  convectively-enhanced dissolution processes in porous media.}
\newblock {\em Advances in Water Resources\/}, 99, 1--14.

\bibitem[\protect\citename{Drignei et~al., }2008]{drignei2008parameter}
Drignei, D., Forest, C.~E., Nychka, D., et~al. (2008).
\newblock \enquote{Parameter estimation for computationally intensive nonlinear
  regression with an application to climate modeling.}
\newblock {\em The Annals of Applied Statistics\/}, 2, 4, 1217--1230.

\bibitem[\protect\citename{Flury and Shephard, }2011]{flury2011bayesian}
Flury, T. and Shephard, N. (2011).
\newblock \enquote{Bayesian inference based only on simulated likelihood:
  particle filter analysis of dynamic economic models.}
\newblock {\em Econometric Theory\/}, 27, 5, 933--956.

\bibitem[\protect\citename{Gattiker et~al., }2016]{gattuq:2016}
Gattiker, J., Myers, K., Williams, B., Higdon, D., Carzolio, M., and Hoegh, A.
  (2016).
\newblock \enquote{Gaussian Process-Based Sensitivity Analysis and Bayesian
  Model Calibration with GPMSA.}
\newblock In {\em Handbook of Uncertainty Quantification\/}, eds. R.~Ghanem,
  D.~Higdon, and H.~Owhadi,  1867--1907. Switzerland: Springer.

\bibitem[\protect\citename{Henderson et~al., }2009]{henderson2009bayesian}
Henderson, D.~A., Boys, R.~J., Krishnan, K.~J., Lawless, C., and Wilkinson,
  D.~J. (2009).
\newblock \enquote{Bayesian emulation and calibration of a stochastic computer
  model of mitochondrial DNA deletions in substantia nigra neurons.}
\newblock {\em Journal of the American Statistical Association\/}, 104, 485,
  76--87.

\bibitem[\protect\citename{Higdon et~al., }2008]{higdon2008cmc}
Higdon, D., Gattiker, J., Williams, B., and Rightley, M. (2008).
\newblock \enquote{{Computer Model Calibration Using High-Dimensional Output}.}
\newblock {\em Journal of the American Statistical Association\/}, 103, 482,
  570--583.

\bibitem[\protect\citename{Higdon et~al., }2004]{higd:kenn:cave:2004}
Higdon, D., Kennedy, M., Cavendish, J., Cafeo, J., and Ryne, R.~D. (2004).
\newblock \enquote{Combining field observations and simulations for calibration
  and prediction.}
\newblock {\em SIAM Journal of Scientific Computing\/}, 26, 448--466.

\bibitem[\protect\citename{Higdon et~al., }2002]{higdon2002space}
Higdon, D. et~al. (2002).
\newblock \enquote{Space and space-time modeling using process convolutions.}
\newblock {\em Quantitative methods for current environmental issues\/}, 3754,
  37--56.

\bibitem[\protect\citename{Hooten and Wikle, }2010]{hooten2010statistical}
Hooten, M.~B. and Wikle, C.~K. (2010).
\newblock \enquote{Statistical agent-based models for discrete spatio-temporal
  systems.}
\newblock {\em Journal of the American Statistical Association\/}, 105, 489,
  236--248.

\bibitem[\protect\citename{Kattwinkel and Reichert,
  }2017]{kattwinkel2017bayesian}
Kattwinkel, M. and Reichert, P. (2017).
\newblock \enquote{Bayesian parameter inference for individual-based models
  using a Particle Markov Chain Monte Carlo method.}
\newblock {\em Environmental Modelling \& Software\/}, 87, 110--119.

\bibitem[\protect\citename{Kennedy and O'{H}agan, }2001]{kenn:ohag:2001}
Kennedy, M. and O'{H}agan, A. (2001).
\newblock \enquote{{B}ayesian calibration of computer models (with
  discussion).}
\newblock {\em Journal of the Royal Statistical Society (Series B)\/}, 68,
  425--464.

\bibitem[\protect\citename{Kleijnen, }2009]{kleijnen2009kriging}
Kleijnen, J.~P. (2009).
\newblock \enquote{Kriging metamodeling in simulation: A review.}
\newblock {\em European journal of operational research\/}, 192, 3, 707--716.

\bibitem[\protect\citename{Lawrence et~al., }2010]{lawrence2010coyote}
Lawrence, E., Heitmann, K., White, M., Higdon, D., Wagner, C., Habib, S., and
  Williams, B. (2010).
\newblock \enquote{The coyote universe. III. simulation suite and precision
  emulator for the nonlinear matter power spectrum.}
\newblock {\em The Astrophysical Journal\/}, 713, 2, 1322.

\bibitem[\protect\citename{Lin and Dunson, }2014]{lin2014bayesian}
Lin, L. and Dunson, D.~B. (2014).
\newblock \enquote{Bayesian monotone regression using Gaussian process
  projection.}
\newblock {\em Biometrika\/}, 101, 2, 303--317.

\bibitem[\protect\citename{Marrel et~al., }2012]{marrel2012global}
Marrel, A., Iooss, B., Da~Veiga, S., and Ribatet, M. (2012).
\newblock \enquote{Global sensitivity analysis of stochastic computer models
  with joint metamodels.}
\newblock {\em Statistics and Computing\/}, 22, 3, 833--847.

\bibitem[\protect\citename{Oakley and O'Hagan, }2004]{oakl:ohag:2004}
Oakley, J. and O'Hagan, A. (2004).
\newblock \enquote{Probabilistic sensitivity analysis of complex models.}
\newblock {\em Journal of the Royal Statistical Society (Series B)\/}, 66,
  751--769.

\bibitem[\protect\citename{Owada et~al., }2016]{owada2016epidemiological}
Owada, K., Eckmanns, T., Kamara, K.-B.~O., and Olu, O.~O. (2016).
\newblock \enquote{Epidemiological data management during an outbreak of Ebola
  virus disease: key issues and observations from Sierra Leone.}
\newblock {\em Frontiers in public health\/}, 4.

\bibitem[\protect\citename{Paulo et~al., }2012]{paulo2012calibration}
Paulo, R., Garc{\'\i}a-Donato, G., and Palomo, J. (2012).
\newblock \enquote{Calibration of computer models with multivariate output.}
\newblock {\em Computational Statistics \& Data Analysis\/}, 56, 12,
  3959--3974.

\bibitem[\protect\citename{Plumlee and Tuo, }2014]{plumlee2014building}
Plumlee, M. and Tuo, R. (2014).
\newblock \enquote{Building accurate emulators for stochastic Simulations via
  quantile Kriging.}
\newblock {\em Technometrics\/}, 56, 4, 466--473.

\bibitem[\protect\citename{Reich et~al., }2012]{reich2012variable}
Reich, B.~J., Kalendra, E., Storlie, C.~B., Bondell, H.~D., and Fuentes, M.
  (2012).
\newblock \enquote{Variable selection for high dimensional Bayesian density
  estimation: application to human exposure simulation.}
\newblock {\em Journal of the Royal Statistical Society: Series C (Applied
  Statistics)\/}, 61, 1, 47--66.

\bibitem[\protect\citename{Sacks et~al., }1989]{sack:welc:mitc:wynn:1989}
Sacks, J., Welch, W.~J., Mitchell, T.~J., and Wynn, H.~P. (1989).
\newblock \enquote{Design and Analysis of Computer Experiments (with
  Discussion).}
\newblock {\em Statistical Science\/}, 4, 409--423.

\bibitem[\protect\citename{Saltelli et~al., }2008]{saltelli2008global}
Saltelli, A., Ratto, M., Andres, T., Campolongo, F., Cariboni, J., Gatelli, D.,
  Saisana, M., and Tarantola, S. (2008).
\newblock {\em Global sensitivity analysis: the primer\/}.
\newblock Wiley.

\bibitem[\protect\citename{Sobol, }2001]{sobol2001global}
Sobol, I.~M. (2001).
\newblock \enquote{Global sensitivity indices for nonlinear mathematical models
  and their Monte Carlo estimates.}
\newblock {\em Mathematics and computers in simulation\/}, 55, 1, 271--280.

\bibitem[\protect\citename{Tabataba et~al., }2017]{tabataba2017framework}
Tabataba, F.~S., Chakraborty, P., Ramakrishnan, N., Venkatramanan, S., Chen,
  J., Lewis, B., and Marathe, M. (2017).
\newblock \enquote{A framework for evaluating epidemic forecasts.}
\newblock {\em BMC infectious diseases\/}, 17, 1, 345.

\bibitem[\protect\citename{Tuo and Jeff~Wu, }2016]{tuo2016theoretical}
Tuo, R. and Jeff~Wu, C. (2016).
\newblock \enquote{A theoretical framework for calibration in computer models:
  parametrization, estimation and convergence properties.}
\newblock {\em SIAM/ASA Journal on Uncertainty Quantification\/}, 4, 1,
  767--795.

\bibitem[\protect\citename{Venkatramanan et~al., }2017]{venkatramanan2017using}
Venkatramanan, S., Lewis, B., Chen, J., Higdon, D., Vullikanti, A., and
  Marathe, M. (2017).
\newblock \enquote{Using data-driven agent-based models for forecasting
  emerging infectious diseases.}
\newblock {\em Epidemics\/}.

\bibitem[\protect\citename{von Storch and Zwiers, }1999]{eof:1999}
von Storch, H. and Zwiers, F.~W. (1999).
\newblock {\em Statistical Analysis in Climate Research\/}.
\newblock New York: Cambridge University Press.

\bibitem[\protect\citename{Wang and Berger, }2016]{wang2016estimating}
Wang, X. and Berger, J.~O. (2016).
\newblock \enquote{Estimating shape constrained functions using Gaussian
  processes.}
\newblock {\em SIAM/ASA Journal on Uncertainty Quantification\/}, 4, 1, 1--25.

\bibitem[\protect\citename{{WHO Ebola Response Team}, }2014]{team2014ebola}
{WHO Ebola Response Team} (2014).
\newblock \enquote{Ebola virus disease in West Africa—the first 9 months of
  the epidemic and forward projections.}
\newblock {\em The New England journal of medicine\/}, 371, 16, 1481.

\bibitem[\protect\citename{Wikle and Hooten, }2016]{wikle2016hierarchical}
Wikle, C.~K. and Hooten, M.~B. (2016).
\newblock \enquote{Hierarchical Agent-Based Spatio-Temporal Dynamic Models for
  Discrete-Valued Data.}
\newblock {\em Handbook of Discrete-Valued Time Series\/}, 349.

\bibitem[\protect\citename{Ye et~al., }2000]{ye2000algorithmic}
Ye, K., Li, W., and Sudjianto, A. (2000).
\newblock \enquote{Algorithmic construction of optimal symmetric Latin
  hypercube designs.}
\newblock {\em Journal of statistical planning and inference\/}, 90, 1,
  145--159.

\end{thebibliography}
\bibliographystyle{jasa}

\newpage

\end{document}